\documentclass[cits]{PoS}

\usepackage{bm}

\DeclareMathAlphabet{\mathcal}{OMS}{cmsy}{m}{n}

\title{Future applications of the Yang--Mills gradient flow in lattice QCD}

\ShortTitle{Yang--Mills gradient flow}

\author{\speaker{M.~L\"uscher}\\
        CERN, Physics Department, 1211 Geneva 23, Switzerland\\
        E-mail: \email{luscher@mail.cern.ch}}

\abstract{
The Yang--Mills gradient flow has many interesting applications
in lattice QCD. In this talk, some recent and possible future
uses of the flow are discussed, emphasizing the underlying
theoretical concepts rather than any computational aspects.
\\[0.5cm]
\rightline{CERN-PH-TH/2013-201}
}

\FullConference{31st International Symposium on Lattice Field Theory - 
                LATTICE 2013\\
		July 29 - August 3, 2013\\
		Mainz, Germany}

\begin{document}



\def\rmd{{\rm d}}
\def\rmD{{\rm D}}
\def\rme{{\rm e}}
\def\rmO{{\rm O}}
\def\rmU{{\rm U}}


\def\rz{{\mathbb R}}
\def\gz{{\mathbb Z}}
\def\nz{{\mathbb N}}
\def\Im{{\rm Im}\,}
\def\Re{{\rm Re}\,}


\def\defeq{\mathrel{\mathop=^{\rm def}}}
\def\proof{\noindent{\sl Proof:}\kern0.6em}
\def\endproof{\hskip0.6em plus0.1em minus0.1em
\setbox0=\null\ht0=5.4pt\dp0=1pt\wd0=5.3pt
\vbox{\hrule height0.8pt
\hbox{\vrule width0.8pt\box0\vrule width0.8pt}
\hrule height0.8pt}}
\def\sfrac#1#2{\hbox{$\frac{#1}{#2}$}}
\def\dual{\mathstrut^*\kern-0.1em}
\def\mod{\;\hbox{\rm mod}\;}
\def\ring{\mathaccent"7017}
\def\slash#1{\setbox2=\hbox{$\displaystyle#1$}%
             \setbox3=\hbox{$\displaystyle/$}%
             #1\kern-0.8\wd2/\kern-1.0\wd3\kern0.8\wd2\kern0.5pt}
\def\lvec#1{\setbox0=\hbox{$#1$}
    \setbox1=\hbox{$\scriptstyle\leftarrow$}
    #1\kern-\wd0\smash{
    \raise\ht0\hbox{$\raise1pt\hbox{$\scriptstyle\leftarrow$}$}}
    \kern-\wd1\kern\wd0}
\def\rvec#1{\setbox0=\hbox{$#1$}
    \setbox1=\hbox{$\scriptstyle\rightarrow$}
    #1\kern-\wd0\smash{
    \raise\ht0\hbox{$\raise1pt\hbox{$\scriptstyle\rightarrow$}$}}
    \kern-\wd1\kern\wd0}
\def\cvec#1{\kern-0.5pt\vec{\kern0.5pt #1}}
\def\boxit#1{\vbox{\hrule height2pt\hbox{\vrule width2pt
    \kern10pt\vbox{\kern10pt#1\kern10pt}\kern10pt\vrule width2pt}
    \hrule height2pt}}
\def\Arw{{\redtwo\bm\Rightarrow}}

\def\wick#1{\setbox2=\hbox{$\displaystyle#1$}
    \setbox3=\null\ht3=3.0pt\dp3=0.0pt\wd3=20.0pt
    #1\kern-\wd2\kern3.0pt\raise10.0pt\vbox{\hrule height0.3pt
    \hbox{\vrule width0.3pt\box3\vrule width0.3pt}}\kern-24.0pt\kern\wd2}
\def\longwick#1{\setbox2=\hbox{$\displaystyle#1$}
    \setbox3=\null\ht3=3.0pt\dp3=0.0pt\wd3=28.0pt
    #1\kern-\wd2\kern3.0pt\raise10.0pt\vbox{\hrule height0.3pt
    \hbox{\vrule width0.3pt\box3\vrule width0.3pt}}\kern-32.0pt\kern\wd2}
\def\verylongwick#1{\setbox2=\hbox{$\displaystyle#1$}
    \setbox3=\null\ht3=3.0pt\dp3=0.0pt\wd3=43.0pt
    #1\kern-\wd2\kern3.0pt\raise10.0pt\vbox{\hrule height0.3pt
    \hbox{\vrule width0.3pt\box3\vrule width0.3pt}}\kern-47.0pt\kern\wd2}


\def\nab#1{{\nabla\kern-1.5pt_{#1}\kern1.0pt}}
\def\nabstar#1{\nabla\kern0.5pt\smash{\raise5.0pt\hbox{$\ast$}}
               \kern-7.5pt_{#1}\kern2.0pt}
\def\drv#1{{\partial_{#1}}}
\def\drvstar#1{\partial\kern1.0pt\smash{\raise4.5pt\hbox{$\ast$}}
               \kern-6.5pt_{#1}\kern2pt}
\def\drvtilde#1#2{{\tilde{\partial}_{#1}^{#2}}}


\def\MeV{{\rm MeV}}
\def\GeV{{\rm GeV}}
\def\TeV{{\rm TeV}}
\def\fm{{\rm fm}}
\def\MSbar{\overline{\rm MS\kern-0.5pt}\kern0.5pt}


\def\euler{\gamma_{\rm E}}


\def\psibar{\overline{\psi}}
\def\chibar{\overline{\chi}}
\def\psitilde{\widetilde{\psi}}
\def\ubar{\bar{u}}
\def\dbar{\bar{d}}
\def\sbar{\bar{s}}
\def\Ubar{V}


\def\dirac#1{\gamma_{#1}}
\def\diracstar#1#2{
    \setbox0=\hbox{$\gamma$}\setbox1=\hbox{$\gamma_{#1}$}
    \gamma_{#1}\kern-\wd1\kern\wd0
    \smash{\raise4.5pt\hbox{$\scriptstyle#2$}}}
\def\dirachat{\hat{\gamma}_5}


\def\SUtwo{{\rm SU(2)}}
\def\SUthree{{\rm SU(3)}}
\def\SUn{{\rm SU}(N)}
\def\tr{{\rm tr}}
\def\Tr{{\rm Tr}}
\def\Ad{{\rm Ad}\,}
\def\Group{{\rm SU}(3)}
\def\Lie{\mathfrak{su}(3)}


\def\Nf{N_{\rm f}}


\def\obs{{\mathcal O}}
\def\Mpi{M_{\pi}}
\def\Fpi{F_{\pi}}
\def\Gpi{G_{\pi}}
\def\Gpit{G_{\pi,t}}
\def\MK{M_K}
\def\gbar{\bar{g}}
\def\Sigt{\Sigma_t}
\def\Zchi{Z_{\chi}}
\def\ZP{Z_P}
\def\cfl{c_{\rm fl}}
\def\mq#1{m_{{\rm q},#1}}
\def\Mq{M_{\rm q}}


\def\ren#1{#1_{\rm R}}
\def\rent#1{#1_{{\rm R},t}}
\def\eps{\epsilon}
\def\vsp{{\vphantom{$a_b$}}}
\def\thicktablerule{\hrule height1pt}
\def\thintablerule{\hrule height0.4pt}

\section{Introduction}

In the past few years,
the Yang--Mills gradient flow turned out to be a
theoretically attractive and powerful tool for
non-perturbative studies of QCD.
A key feature of the flow is certainly the fact that
local fields constructed at positive flow
time renormalize in a simple way, however complicated
they may be \cite{WilsonFlow,RenFlow,ChFlow}. 
Correlation functions of such fields calculated in lattice
QCD therefore have a well-defined continuum limit
and thus provide interesting probes of the universal
properties of the theory.

Apart from a brief recap of some basic facts and equations, the subject 
will not be reviewed in this talk. Instead the focus is on applications 
of the gradient flow which have not or just barely been considered so far.
The computation of the chiral condensate via
the quark density at positive flow times is one of them,
and its discussion here partly serves as an introduction 
to important concepts such
as the small flow-time expansion of local fields.
There are potentially very interesting uses of the latter in
studies of QCD at non-zero temperatures and in computations of 
electro-weak transition matrix elements, for example.

Another topic covered is the application of the gradient flow 
in the context of renormalization and the continuum limit.
While the flow itself is not a renormalization group transformation,
non-perturbative renormalization can be based
on observables at positive flow times. 
Calculations of the scale evolution of couplings and fields,
using step scaling, as well as systematic constructions 
of coarse-grid actions through a matching procedure are 
likely to profit from the use of such observables, since their 
expectation values can often be accurately computed 
with a modest effort.

\section{The Yang--Mills gradient flow in QCD}

\subsection{Flow equations}

The Yang--Mills gradient flow $B_{\mu}(t,x)$, $t\geq0$,
of $\SUthree$ gauge potentials is defined by the flow equation
\begin{eqnarray}
  &&\hspace{-0.5em}\partial_tB_{\mu}=D_{\nu}G_{\nu\mu},
  \label{GFloweq}\\[2.0ex]
  &&\hspace{-0.5em}G_{\mu\nu}=\partial_{\mu}B_{\nu}-\partial_{\nu}B_{\mu}
  +[B_{\mu},B_{\nu}],
  \qquad
  D_{\mu}=\partial_{\mu}+[B_{\mu},\,\cdot\,],
\end{eqnarray}
and the initial condition $B_{\mu}(0,x)=A_{\mu}(x)$,
where $A_{\mu}(x)$ denotes the fundamental gauge field integrated
over in the QCD functional integral. Since the flow equation
is of first order in the derivatives with respect to the flow time $t$, the 
gauge potentials $B_{\mu}(t,x)$ are uniquely determined by their initial
value at $t=0$ and are thus well-defined functions of
the fundamental gauge field.

Starting from the fundamental quark field $\chi(0,x)=\psi(x)$
at flow time zero,
an associated flow $\chi(t,x)$ of quark fields may be similarly defined
through the equations
\begin{eqnarray}
  &&\hspace{-0.5em}\partial_t\chi=\Delta\chi,
  \label{FFloweq}\\[2.0ex]
  &&\hspace{-0.5em}\Delta=D_{\mu}D_{\mu},
  \qquad
  D_{\mu}=\partial_{\mu}+B_{\mu}.
\end{eqnarray}
Clearly, $\chi(t,x)$ is a uniquely determined linear function of the 
fundamental quark field and a non-linear function of the gauge
potential $B(s,y)$, $0\leq s\leq t$, and therefore of the 
fundamental gauge field.
In the flow equation (\ref{FFloweq}),
the gauge-covariant Laplacian $\Delta$ could be replaced by the square of 
the Dirac operator, but while this choice would
be mathematically appealing, it does
not appear to offer any particular advantages 
in the present context.

It goes without saying that a regularization is required in the
quantized theory for the gradient flow and the associated quark
flow to be well defined. 
Dimensional regularization is a possible choice in perturbation theory
and it is straightforward to set up both the gauge and the quark 
flow on the lattice
\cite{WilsonFlow,ChFlow}. For simplicity, I however 
continue to use a continuum notation and implicitly 
assume a lattice regularization.

\subsection{Smoothing property}

In the direction of increasing flow time, 
the evolution of the time-dependent gauge and quark fields 
tends to have a smoothing effect.
It is possible to show this by expanding
the solution of the flow equations
in powers of the fundamental fields.
At the leading order of the expansion,
\begin{eqnarray}
  &&\hspace{-0.5em}B_{\mu}(t,x)=
  \int\rmd^4y\,K_t(x-y)A_{\mu}(y)+
  \hbox{gauge \& non-linear terms},
  \label{GSmooth}
  \\[2.0ex]
  &&\hspace{-0.5em}\chi(t,x)=\int\rmd^4y\,K_t(x-y)\psi(y)+\ldots,
  \qquad
  K_t(z)\propto\exp\Bigl\{-\frac{z^2}{4t}\Bigr\},
  \label{FSmooth}
\end{eqnarray}  
the fields are then seen to be smoothed with a Gaussian kernel over 
a spherical range of distances with radius roughly equal to 
$\sqrt{8t}$. The smoothing property is unchanged at
higher orders and is also observed in numerical studies, the
smoothing range being practically the same as the one given by
the leading-order formulae (\ref{GSmooth}),(\ref{FSmooth}).

The Yang--Mills gradient flow was probably first considered
in mathematics by Atiyah and Bott \cite{AtiyahBott}
in their seminal work on the Morse theory of the space of gauge fields
on two-dimensional manifolds.
In lattice QCD, field smoothing techniques are 
in use since many years. Repeated ``stout link smearing'' 
\cite{Stout} in fact amounts to a numerical integration of
the flow equation (\ref{GFloweq}) using an Euler
scheme, while the source-smearing operations proposed in 
refs.~\cite{Guesken,AlexandrouEtAl} solve the quark flow
equation (\ref{FFloweq}) on an
equal-time hyperplane, albeit without
simultaneously evolving the gauge field.

\subsection{Observables}

\begin{figure}
\centering
\includegraphics[height=.2\textwidth,clip]{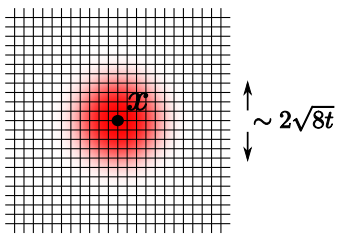}
\caption{Local fields $\obs_t(x)$ constructed at flow time $t>0$ 
depend on the fundamental field variables in a region of space-time
approximately $2\sqrt{8t}$ wide (red area).
Further away from the point $x$, the sensitivity to the basic fields
decreases like a Gaussian and very rapidly becomes totally
negligible.
}
\label{fig1}
\end{figure}

Gauge-invariant composite fields constructed at positive flow time $t$,
such as
\begin{eqnarray}
  &&\hspace{-0.5em}E_t=-\sfrac{1}{2}\tr\{G_{\mu\nu}G_{\mu\nu}\},
  \label{Et}
  \\[2.0ex]
  &&\hspace{-0.5em}P_t^{rs}=\chibar_r\dirac{5}\chi_s,
  \qquad
  S_t^{rs}=\chibar_r\chi_s,
  \qquad
  \hbox{$r,s$: flavour labels},
  \label{PtSt}
\end{eqnarray}
are well-defined functions of the fundamental field variables and 
are therefore possible observables 
like the ordinary composite fields. In particular,
one may be interested in their correlation functions 
and, more generally, the QCD expectation values of any product
of local fields at zero and non-zero flow times.
Clearly, an observable 
$\obs_t(x)$ that is locally constructed at some flow time $t>0$ 
is not a local expressions in the fundamental fields, but
essentially only depends on the field variables
in a spherical region of space-time centered at $x$,
which I will refer to as the footprint of the observable
(see fig.~\ref{fig1}).

The smoothing property of the gradient flow and the associated
quark flow implies that correlation functions of fields at
non-zero flow times have no short-distance singularities.
Renormalization is nevertheless required, but turns out to 
be extremely simple. Explicitly, if $\obs_t(x)$ is a 
bare, gauge-invariant composite field at flow time $t>0$ of
degree $n$ and $\bar{n}$ in the quark and antiquark fields, the 
renormalized field is given by
\begin{equation}
  \rent{\obs}=(\Zchi)^{\frac{1}{2}(n+\bar{n})}\obs_t, 
  \label{RenObs}
\end{equation}
where the renormalization constant $\Zchi$ is independent of $t$.
In particular, the field (\ref{Et}) 
does not require renormalization and the 
chiral densities (\ref{PtSt}) renormalize with the same factor
$\Zchi$.

The proof of these statements \cite{RenFlow,ChFlow}
is based on an exact representation of 
the correlation functions through a local field theory in 
4+1 dimensions, the extra dimension being the flow time.
Zinn--Justin and Zwanziger \cite{ZinnZwanziger} introduced
the representation many years ago in their work on the 
renormalization of the Langevin equation. In the pure gauge theory,
the latter actually coincides with the flow equation 
(\ref{GFloweq}) except for the fact that it includes  
a noise term, which complicates the situation and requires
a renormalization of the Langevin time, for example.

\section{Chiral condensate}

In lattice QCD, the expectation value of the
scalar density $\ubar u+\dbar d$ of the up and down quarks 
diverges like the second or
third inverse power of the lattice spacing when the continuum limit 
is taken. 
The divergent terms are proportional to the light-quark
masses if the lattice theory preserves chiral symmetry,
but also in these cases their subtraction tends to give rise 
to important significance losses or even some conceptual issues.
Using the gradient flow,
this problem can now be elegantly bypassed \cite{ChFlow}.

\subsection{Flow-time dependent condensate}

Since the flow equations are chirally invariant,
the quark field at non-zero flow times, $\chi(t,x)$,
transforms in the same way as the fundamental field $\psi(x)$
under global chiral rotations. 
In particular, the light-quark chiral densities
\begin{equation}
  S^{rs}_t\pm P^{rs}_t,\qquad r,s\in\{u,d\},
  \label{StPt}
\end{equation}
transform according to the $(\frac{1}{2},\frac{1}{2})$ representation
of the $\SUtwo_{\rm L}\times\SUtwo_{\rm R}$ chiral symmetry group
that acts on the up and down quark fields.
The ``time-dependent condensate''
\begin{equation}
  \Sigt=-\sfrac{1}{2}\langle S^{uu}_t+S^{dd}_t\rangle
  \label{Sigt}
\end{equation}
is therefore an order parameter for the spontaneous breaking of 
this symmetry.

From a technical point of view, the condensate (\ref{Sigt}) is 
an attractive quantity, because no power-divergent terms
need to be subtracted, when the continuum limit is taken, and
only multiplicative renormalization is required
according to eq.~(\ref{RenObs}).
Moreover, calculations of $\Sigt$ through numerical simulation
are straightforward and quickly yield accurate results.

\subsection{Relation to $\bm\Sigma$}

The time-dependent condensate $\Sigt$ can be directly used
for studies of spontaneous chiral symmetry breaking at non-zero
temperatures, for example, but one may nevertheless 
be interested in its relation to the standard chiral condensate
$\Sigma$.

\begin{figure}
\centering
\includegraphics[height=.2\textwidth,clip]{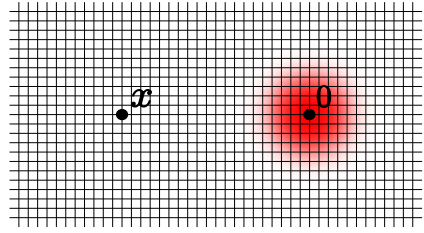}
\caption{
As long as $t$ is strictly positive, the pseudo-scalar two-point function 
in the PCAC relation (\protect\ref{PCAC}) 
has no short-distance singularities.
Moreover, at distances $|x|\gg\sqrt{8t}$,
the transfer matrix formalism may be invoked to show 
that the correlation function falls off proportionally to 
$\rme^{-\Mpi|x|}$.
}
\label{fig2}
\end{figure}

The link between the two condensates is 
provided by the PCAC relation. When probed by
the pseudo-scalar density $P_t^{du}$
at flow time $t$ and integrated
over space-time, the relation implies
\begin{equation}
  \Sigt=-\frac{\Mpi^2\Fpi}{2\Gpi}
  \int\rmd^4x\,\langle P^{ud}(x)P_t^{du}(0)\rangle,
  \label{PCAC}
\end{equation}
where $\Mpi$, $\Fpi$ and $\Gpi$ denote the pion mass, the pion decay constant
and the vacuum-to-pion matrix element of the density $P^{ud}$
(see fig.~\ref{fig2}). At large distances,
the two-point function in this equation falls off exponentially
and another constant, $\Gpit$, may be defined
through the asymptotic formula
\begin{equation}
  \int\rmd^3\vec{x}\,\langle P^{ud}(x)P_t^{du}(0)\rangle
  \mathrel{\mathop\sim_{x_0\to\infty}}
  -\frac{\Gpi\Gpit}{\Mpi}\rme^{-\Mpi x_0}.
  \label{Gpit}
\end{equation}
Now when the up and down quark masses $m_u,m_d$ are sent to zero,
the integral in eq.~(\ref{PCAC}) diverges at large $x$
and a few lines of algebra then show that
\begin{equation}
  \Sigma=
  \lim_{m_u=m_d\to0}\Sigt\frac{\Gpi}{\Gpit}.
  \label{Sig2Sigt}
\end{equation}
In the SU(2) chiral limit, the time-dependent chiral condensate
$\Sigt$ thus coincides with the standard condensate $\Sigma$
up to a computable proportionality constant.

\subsection{Chiral perturbation theory}

How exactly
the chiral limit (\ref{Sig2Sigt}) is reached
can be predicted using chiral perturbation theory.
The important point to note is that
local fields at non-zero flow time with a footprint much smaller 
than the Compton wave length of the pion 
are indistinguishable from strictly local fields
when only their matrix elements between low-energy pion
states are considered.
In the effective chiral theory, and if
\begin{equation}
  8t\Mpi^2\ll1,
\end{equation}
the time-dependent densities (\ref{StPt}) are 
therefore represented by the same fields as
the ordinary chiral densities, except for the fact that the 
coefficients in these expressions are not the same.

To one-loop order of $\SUtwo$ chiral perturbation theory 
\cite{GasserLeutwyler}, the formula
\begin{equation}
  \Sigt\frac{\Gpi}{\Gpit}=\Sigma\left\{1-\frac{3\Mpi^2}{32\pi^2\Fpi^2}
  \ln(\Mpi^2/\Lambda_{t}^2)+\ldots\right\}
  \label{ChSigt}
\end{equation}
is then obtained, where
\begin{equation}
  \bar{l}_{t}=
  \left.\ln(\Lambda_{t}^2/M^2)\right|_{M=139.6\,\MeV}
  \label{LECt}
\end{equation}
is a new low-energy effective constant.
While the form of the expansion (\ref{ChSigt}) looks familiar, the 
constant $\bar{l}_t$ 
is unrelated to the well-known low-energy
constants \cite{GasserLeutwyler} and moreover depends on the flow time $t$.

\subsection{Numerical experiment}

As already mentioned, the time-dependent condensate is easily
computed through numerical simulation of the lattice theory.
For illustration, the results obtained in a first study of this
kind \cite{ChFlow} are now briefly recalled.
In this run (labeled $I_1$ in ref.~\cite{openQCDII}),
the O($a$)-improved lattice 
theory with 2+1 flavours of Wilson quarks was simulated
on a $64\times32^3$ lattice with open boundary conditions in time
\cite{openQCDI}. At the chosen point in parameter space,
the lattice spacing is about $0.09$ fm \cite{AokiEtAlI,AokiEtAlII}
and the pion and kaon masses are 
approximately equal to $203$ and $520$ MeV \cite{openQCDII},
respectively.

\begin{table}
\centering
\newdimen\digitwidth
\setbox0=\hbox{\rm 0}
\digitwidth=\wd0
\catcode`@=\active
\def@{\kern\digitwidth}
\tabcolsep0.50cm
\begin{tabular}{ccc}
$\displaystyle\sqrt{8t}$ [fm] &
$\displaystyle a^3\ZP^{-1}\Sigt$ &
$\displaystyle a^3\ZP^{-1}\Sigt\frac{\Gpi}{\Gpit}$\\
\noalign{\vskip0.5ex}
\hline\hline
\noalign{\vskip0.5ex}
   $0.4$ &
   $0.0006277(95)$ &
   $0.003962(61)$ \\
   $0.5$ &
   $0.0004251(58)$ &
   $0.003872(55)$ \\
   $0.6$ &
   $0.0002911(36)$ &
   $0.003785(51)$ \\
   $0.7$ &
   $0.0002017(23)$ &
   $0.003711(48)$ \\
\noalign{\vskip0.5ex}
\hline\hline
\end{tabular}
\caption{
Simulation results for the unrenormalized time-dependent condensate 
and the ratio of constants on the 
right of eq.~(\protect\ref{Sig2Sigt}). The values are given in lattice units
and the quoted errors are statistical only. 
}
\label{tab1}
\end{table}

Using a representative ensemble of $150$ 
gauge configurations, the values obtained for the time-dependent
condensate do indeed have small statistical errors (see table~\ref{tab1}).
As a function of the flow time $t$, the condensate decreases monotonically,
but this trend
is practically compensated by the ratio of 
$G$-factors multiplying the condensate in eq.~(\ref{Sig2Sigt}).
The chiral correction 
(\ref{ChSigt}) thus appears to be small
at the quark masses and flow times considered.
Moreover, insertion of the lattice spacing $a=0.08995(40)$ fm 
and the renormalization constant $\ZP=0.5800(47)$ 
of the pseudo-scalar density determined by the 
PACS-CS collaboration \cite{AokiEtAlII,AokiEtAlIV,Taniguchi} yields
\begin{equation}
  \Sigt\frac{\Gpi}{\Gpit}=[287(2)\,\MeV]^3
  \quad\hbox{at}\quad
  \sqrt{8t}=0.5\,\fm
\end{equation}
for the renormalized ratio in the $\MSbar$ scheme at $2$ GeV, 
i.e.~a value well within the range 
expected from previous calculations of the condensate $\Sigma$.

\section{Small flow-time expansion}

In all these calculations, no power-divergent terms were encountered
nor were there any large contributions that had to be subtracted or 
extrapolated away.
Through the small flow-time expansion,
some further insight can be gained into how exactly 
the power divergences are avoided.

\subsection{General form of the expansion}

Let $\obs_t(x)$ be a gauge-invariant local field 
in the continuum theory at flow time $t>0$.
Since its footprint decreases proportionally to $\sqrt{8t}$,
the field looks more and more like a strictly local field when
$t$ is taken to zero.
Eventually, $\obs_t(x)$ can be represented through
an asymptotic expansion \cite{RenFlow}
\begin{equation}
  \obs_t(x)\mathrel{\mathop\sim_{t\to0}}
  \sum_kc_k(t)\phi_k(x)
  \label{SmallExp}
\end{equation}
in a series of local fields $\phi_k(x)$ at 
vanishing flow time with time-dependent coefficients $c_k(t)$.
The expansion holds 
when inserted in correlation functions at
non-zero distances and if all fields are properly renormalized.
Clearly, for the expansion to be completely well defined,
the chosen renormalization conditions must fix both 
the normalization and mixing of the fields $\phi_k(x)$.

The expansion coefficients $c_k(t)$ satisfy a renormalization
group equation that determines their asymptotic behaviour
\begin{equation}
  c_k(t)\mathrel{\mathop\propto_{t\to0}}
  t^{\frac{1}{2}(d_k-d_O)}\gbar^{\nu_k}\{1+\rmO(\gbar^2)\}
  \label{ckt}
\end{equation}
at small flow times. In this equation, $d_k$ and $d_O$ are the
engineering dimensions of the fields $\phi_k(x)$ and $\obs_t(x)$,
the exponents $\nu_k$ are determined by the one-loop coefficients
of their anomalous dimensions and $\gbar$ denotes the running 
coupling (in any scheme) at momentum $(8t)^{-1/2}$
\cite{RenFlow}. The fields $\phi_k(x)$ with the lowest dimension
thus dominate in the limit $t\to0$, all other terms being
suppressed by powers of $t$.

\subsection{Example: expansion of the chiral densities 
(\protect\ref{StPt})}

The asymptotic expressions for the expansion coefficients
$c_k(t)$ simplify considerably if the
renormalization-group-invariant (RGI) normalization convention
is adopted for the fields and
the quark mass matrix $M$
(the RGI convention is described
in sect.~2.2 of ref.~\cite{WeiszHouches}, for example).
In the case of the densities (\ref{StPt}), the
small flow-time expansion then assumes the form
\begin{eqnarray}
  &&\hspace{-0.5em}
  S_t^{rs}(x)=
  c_0(t)M^{rs}+c_1(t)\tr\{M^2\}M^{rs}+c_2(t)(M^3)^{rs}
  +c_3(t)S^{rs}(x)+\rmO(t),
  \label{StExp}\\[2.0ex]
  &&\hspace{-0.5em}
  P_t^{rs}(x)=c_3(t)P^{rs}(x)+\rmO(t),
  \label{PtExp}\\[2.0ex]
  &&\hspace{-0.5em}
  c_0(t)=-\frac{3}{8\pi^2t}\{1+\rmO(\gbar^2)\},
  \qquad
  c_3(t)=(2b_0\gbar^2)^{-8/9}\{1+\rmO(\gbar^2)\},
\end{eqnarray}
where $b_0$ is the one-loop coefficient of the QCD
$\beta$-function. The structure of these expansions is partly
dictated by chiral symmetry. Moreover, the asymptotic
relation
\begin{equation}
  c_3(t)=\frac{\Gpit}{\Gpi}+\rmO(t)
  \label{cPtGpi}
\end{equation}
holds,
as can easily be shown by inserting eq.~(\ref{PtExp}) in 
eq.~(\ref{Gpit}).

When $t$ is sent to zero, the first term in the expansion 
(\ref{StExp}) of the scalar density blows up like $1/t$
and eventually makes the dominant contribution to the 
time-dependent condensate $\Sigt$. Clearly, for the 
ratio $\Sigt\Gpit/\Gpi$ to be close to $\Sigma$,
the flow time and the quark masses must be such that
this singular term is much smaller than $\Sigt$. 
Some rough estimates suggest that 
this condition is indeed satisfied 
in the numerical experiment reported earlier.

\subsection{A broader perspective}

The small flow-time expansion may also be used
to represent any gauge-invariant local field $\phi(x)$ 
at vanishing flow time through an asymptotic 
series of local fields at some positive flow time $t$. 
In the simplest cases,
where the fields are uniquely characterized by their dimension and
symmetry properties, the representation assumes the form
\begin{equation}
  \phi(x)=c(t)\obs_t(x)+\rmO(t)
  \label{OpRep}
\end{equation}
with a coefficient $c(t)$ that varies at most logarithmically as 
$t$ goes to zero. 
Fields where such a representation may conceivably be
worth considering include the energy-momentum tensor \cite{Suzuki}
and the various parts of the effective electro-weak Hamiltonian.

The use of local fields at non-zero
flow time is attractive, because their renormalization
and O($a$)-improvement \cite{ChFlow} is very much simpler than
the one of the ordinary local fields \cite{SW,SFimp,ImpNonD}.
As already noted, the renormalization is always multiplicative
by a power of the quark-field renormalization constant $\Zchi$.
The O($a$)-improvement, on the other hand,
is achieved by including a mass correction
\begin{eqnarray}
  &&\hspace{-0.5em}
  \Zchi\to\Zchi(1+b_{\chi}a\mq{r}+\bar{b}_{\chi}
  \tr\,\Mq),
  \label{OaZchi}\\[2.0ex]
  &&\hspace{-0.5em}
  \Mq={\rm diag}(\mq{u},\mq{d},\ldots):\hspace{0.25em}
  \hbox{bare quark mass matrix},
\end{eqnarray}
in the renormalization factor of the quark field with flavour index $r$
and by adding a contact term 
\begin{equation}
  \longwick{\chi(t,x)\chibar(s,y)}\to
  a^8\sum_{v,w}K(t,x;0,v)\{S(v,w)-a\cfl\delta(v-w)
  \}K(s,y;0,w)^{\dagger}
  \label{OaCfl}
\end{equation}
to the quark propagator $S(v,w)$ in the contractions of 
the quark fields at positive flow times.
The kernel $K(t,x;s,y)$ in eq.~(\ref{OaCfl}) denotes the smoothing
factor determined by the quark flow equation and $\cfl$ is a
new improvement coefficient that can be determined 
non-perturbatively from a combination of pseudo-scalar
correlation functions \cite{ChFlow}. In particular,
O($a$)-improvement does not require subtractions of higher-dimensional
fields.

For field representations such as (\ref{OpRep})
to work out in practice, one must however be able
to accurately determine the coefficients multiplying 
the time-dependent fields.
In perturbation theory, 
the coefficients of the fields of dimension $4$ 
in the expansion
of the energy-momentum tensor in the pure gauge theory 
were recently computed to one-loop order \cite{Suzuki}.
Non-perturbative calculations of some of these coefficients
may also be possible using the Ward identities
that derive from the conservation of the energy-momentum tensor
in the continuum theory \cite{DelDebbioI}. 
Another more widely applicable method is to extract
the coefficients from the asymptotic relations
obtained by inserting the field representation in suitable correlation
functions, as in the case of 
the coefficient (\ref{cPtGpi}), but
whether step-scaling strategies can be used in this
context is not clear at present.

Once the coefficients are known, calculations of hadronic 
matrix elements will, in general, require a ``scaling window'',
where
\begin{equation}
  a\ll\sqrt{8t}\ll\hbox{relevant low-energy scales},
  \label{window}
\end{equation}  
as otherwise the lattice effects and the systematic errors
deriving from the neglected terms in the field representation 
are not guaranteed to be small. In this respect,
the computation of the chiral condensate along
the lines described in this talk is a somewhat special case, 
because the contributions of the neglected terms
either vanish in the chiral limit or cancel in 
the product $\Sigt\Gpi/\Gpit$.

\section{Wilson's renormalization group revisited}

In 1979 Wilson gave a memorable lecture at Carg\`ese, where he 
proposed to use the ``blockspin renormalization group'' 
for non-perturbative renormalization and 
the construction of coarse-grid actions
in non-Abelian gauge theories \cite{WilsonCargese}.
An assumption made at the time was that 
the expectation values of iteratively blocked Wilson loops
have a continuum limit. Wilson was well aware of this, but
remarked that a proof of scaling
in lattice perturbation theory would be 
``rather complicated to carry out''.

We may now consider replacing the link blocking by the gradient
flow and to combine the latter with
step scaling \cite{StepScaling}, which may be regarded as a particular 
incarnation of the renormalization group.
For simplicity, the quark masses are set to zero in the following,
but the extension of the discussion to the general case is 
straightforward and is
of some interest, since mass-independent renormalization 
schemes are inappropriate for the charm and the heavier quarks.

\subsection{Step scaling}

\begin{figure}
\centering
\includegraphics[width=.75\textwidth,clip]{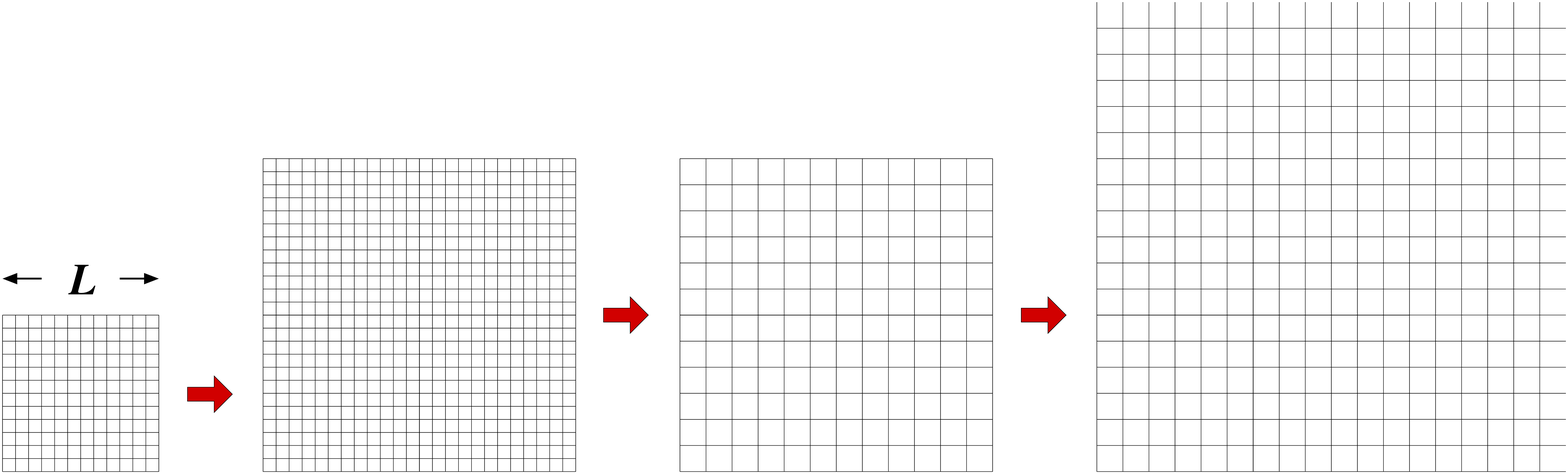}
\caption{
In step-scaling calculations, the scale evolution amounts to 
changing the lattice size $L$ at fixed 
lattice spacing $a$ (i.e.~fixed bare coupling)
by, say, a factor $2$.
Then follows a renormalization step, where 
the running coupling $\gbar^2(L)$ (and thus $L$ in physical units) is held
fixed, while $a$ is increased by a factor $2$. After that
the process continues with another scale evolution step,
and so on.
}
\label{fig3}
\end{figure}

Step scaling is a finite-size scaling method, where the
gauge coupling and the renormalization-scale-dependent 
fields run with the lattice size $L$ (see fig.~\ref{fig3}).
The precise choice of the normalization conditions 
for the coupling and fields
is only of practical importance in this context.
In particular, using the gradient flow, a possible definition of the 
running coupling is \cite{WilsonFlow,NogradiEtAl,FritzschRamos,Ramos}
\begin{equation}
  \gbar^2(L)=\hbox{constant}
  \times\left\{t^2\langle E_t\rangle\right\}_{\sqrt{8t}=\frac{1}{3}L}.
  \label{Gbar}
\end{equation}
As discussed before, the field $E_t$ does not require renormalization.
Moreover, since the flow time $t$ in eq.~(\ref{Gbar}) 
is scaled proportionally to $L^2$,
the coupling depends on a single external scale
and thus runs with $L$ as it should
(the proportionality constant is a parameter of the renormalization
scheme, which may in principle be set to any value
\cite{FritzschRamos}).

The coupling (\ref{Gbar}) has some good technical properties
that should allow step-scaling studies to be performed 
with unprecedented precision. In particular, its computation
in numerical lattice QCD does not require any fits or 
extrapolations to be performed, nor does its variance 
increase significantly towards the large-volume regime
of the theory  \cite{Ramos}, where contact is made 
with the non-perturbative low-energy scales of the theory.

\subsection{Construction of improved actions}

Step scaling is usually combined with an extrapolation to the continuum limit
of the so-called step-scaling functions \cite{StepScaling}. 
The calculation then effectively
solves the non-perturbative renormalization problem
in the continuum theory. 
When constructing improved lattice actions, 
the goal is a different one, namely to reduce the lattice effects
as much as possible after having properly taking into account 
the parameter and field renormalization. 

For practical reasons, improved lattice theories 
may not be too complicated. Their construction
must therefore start from a suitable ansatz for the action and
the improved fields with a reasonably small number of parameters.
The tuning of the parameters of the improved theory
then proceeds by matching lattices with different spacings
as in fig.~4 \cite{WilsonCargese}.

To be able to do this, a sufficiently large set of accurately 
computable renormalized quantities is needed. So far this requirement
was difficult to meet, but there is now new hope
that progress can be made using observables based 
on the Yang--Mills gradient flow and the associated quark flow.
The expectation values of the fields
\begin{eqnarray}
  &&\hspace{-0.5cm}
  \tr\{G_{\mu\nu}G_{\mu\nu}\},\hspace{1.0em}
  \chibar\chi,\hspace{1.0em}
  \chibar\sigma_{\mu\nu}G_{\mu\nu}\chi,
  \\[2.0ex]
  &&\hspace{-0.5cm}  
  (\chibar\,\Gamma \chi)(\chibar\,\Gamma \chi),\quad
  \Gamma\in\{1,\dirac{5},\dirac{\mu},\dirac{\mu}\dirac{5},\sigma_{\mu\nu}\},
\end{eqnarray}
are simple examples of such quantities, and there is obviously a wide range
of further observables, local and non-local ones, that may serve the 
purpose. Moreover, all of them can be considered
at several flow times $t$.
Being able to probe the theory in many
different channels is important in order exclude 
a situation, where the parameter tuning removes some small lattice effects 
in one channel but produces large effects elsewhere.

\begin{figure}
\centering
\includegraphics[width=.64\textwidth,clip]{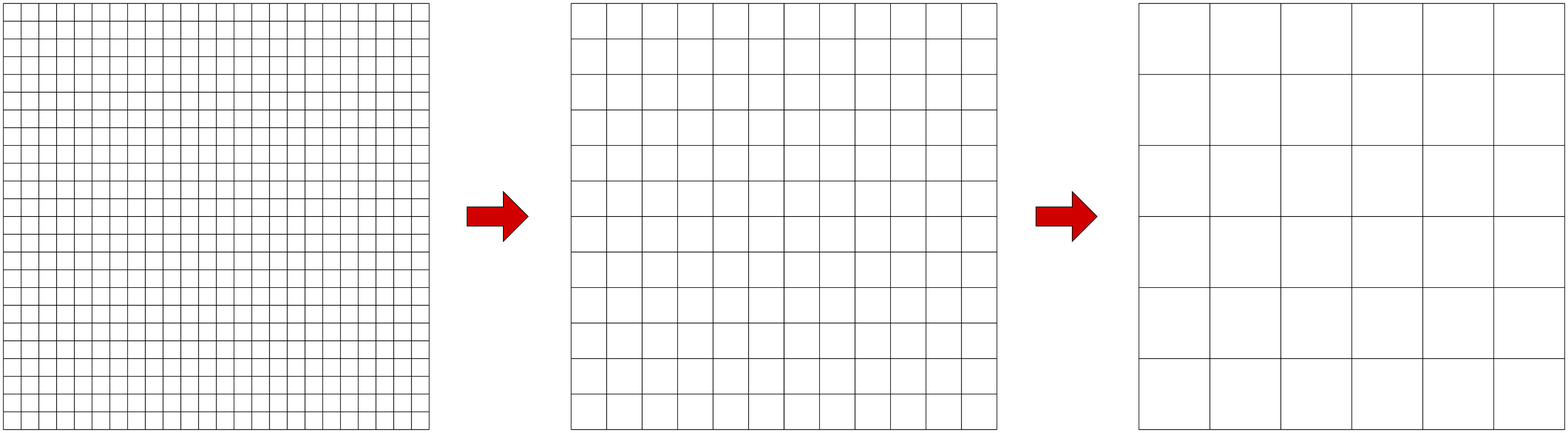}
\caption{
At fixed coupling $\gbar^2(L)$,
renormalized quantities are independent of the lattice spacing
up to lattice effects. Improved actions and fields
can therefore be tuned by simulating sequences of matched lattices
and by minimizing the differences in the measured values of 
a suitable set of such quantities. 
}
\label{fig4}
\end{figure}

\section{Concluding remarks}

Some of the ideas discussed in this talk clearly
need to be further developed, through both analytical and numerical
studies, before their viability can be assessed.
The Yang--Mills gradient flow and its extension
to the quark fields however stand on solid theoretical ground
and there are already a few important applications 
of the flow, which are known to work out in practice.

The fact that the power-divergent mixing of the scalar quark 
density with the unit field can be bypassed by going to 
non-zero flow time is intriguing and suggests that the same
may perhaps be possible in other cases of field mixing as well. 
Computations of electro-weak transition matrix elements,
for example, might profit from such a change of strategy,
but the application of the Yang--Mills gradient flow 
in this context probably belongs to the more distant future.

\acknowledgments

In the course of the preparation of this talk, I have had
many interesting and stimulating discussions on the subject
with Agostino Patella, Roberto Petronzio and Hiroshi Suzuki. 
I also wish to thank Stefan Schaefer and Peter Weisz
for a pleasant and fruitful collaboration in the past few years.
All numerical simulations reported here were performed on a
dedicated PC cluster at CERN. I am grateful
to the CERN management for providing the required funds
and to the CERN IT Department for technical support.

\end{document}